\documentclass[12pt, onecolumn]{IEEEtran}
\usepackage{amsfonts,amssymb, mathtools, amsthm, mathrsfs, latexsym, bbm, bm}
\usepackage[mathscr]{euscript}
\usepackage[T1]{fontenc}
\usepackage[utf8]{inputenc}
\usepackage{lmodern}
\usepackage{newtxtext}
\usepackage{cite, url, hyperref}
\usepackage{url}
\usepackage{setspace}
\def\firstaux#1#2\relax{\lettrine{#1}{#2}}

\usepackage{xcolor}
\newcommand{\mA}{\mathcal{A}}

\usepackage{algorithm}
\usepackage{algpseudocode}

\newcommand{\E}{\mathbbm{E}}

\usepackage{float}

\title{  Angle-of-Arrival Estimation of Narrow Gaussian Beams for Mobile FSO Platforms}
\author{Ming-Cheng~Tsai, \IEEEmembership{Graduate Student~Member,~IEEE}, Muhammad~Salman~Bashir,~\IEEEmembership{Senior~Member,~IEEE}, and Mohamed-Slim~Alouini,~\IEEEmembership{Fellow,~IEEE}
\thanks{M.~-C.~Tsai and M.~-S.~Alouini  are with the King Abdullah University of Science and Technology (KAUST), Thuwal 23955-6900, Kingdom of Saudi Arabia.  e-mail: (mingcheng.tsai@kaust.edu.sa, slim.alouini@kaust.edu.sa). Muhammad Salman Bashir is with the University of Huddersfield, Huddersfield, United Kingdom. e-mail: m.bashir@hud.ac.uk.} 
}

\begin{document}

\maketitle
\begin{abstract}

Due to the narrow beamwidths of laser Gaussian beams, accurate tracking of laser beam's angle-of-arrival is an important problem in mobile free-space optical communications. In most optical receivers today, fine tracking of angle-of-arrival involves estimating the location of the focused beam spot projected onto a focal plane array. However, for very thin Gaussian beams, both the location as well as the energy of the spot varies considerably with the variation of angle-of-arrival. In this study, we have analyzed the relationship between the angle-of-arrival and the energy of laser spot on the focal plane. We then exploited this relationship to enhance the angle-of-arrival estimation performance of our proposed receiver that takes into account both the location as well as the energy of the laser spot while estimating the angle-of-arrival. The derived Cram\'er-Rao bounds indicate that the system performance can be enhanced significantly  for narrow Gaussian beams when both the spot location and energy are exploited for angle-of-arrival estimation.
\end{abstract}

\begin{IEEEkeywords}
Angle-of-arrival, Cram\'er-Rao bound, detector array, estimation, free-space optics, Gaussian beams.
\end{IEEEkeywords}

\section{Introduction}
Due to the availability of large unregulated spectrum in the optical domain of electromagnetic waves, free-space optics (FSO)---also known as optical wireless or laser communications---is an important candidate for supporting high data-rates in the sixth generation (6G) and beyond terrestrial and non-terrestrial wireless networks \cite{Trichili:JOSAB:20}. FSO has been deployed successfully in non-terrestrial networks such as high speed communications between satellites in low-Earth orbit (LEO), medium-Earth orbit (MEO), geosynchronous-equatorial orbit (GEO)  and high-earth orbit (HEO). FSO has also found considerable use in hybrid space-terrestrial networks \cite{Hemmati:JPROC:11}. The National Aeronautics and Space Administration (NASA) demonstrated a downlink data rate up to 622 Mbps from the Moon to the Earth in the Aerospace Corporation’s Optical Communication and Sensor Demonstration (OCSD) in 2017 \cite{Goorjian:SPIE:21}. Additionally, NASA's TeraByte InfraRed Delivery (TBIRD) system promises data rate transfers up to 200 Gbps  from a CubeSat in  LEO to a ground station using laser communications \cite{Chang:MIT:19, Goorjian:SPIE:21}.

 The angular beamwidth $\phi$ of laser Gaussian beam (in the paraxial case) is proportional to $ \frac{\lambda}{\pi}$ \cite{svelto:Springer:13}. This implies that it is not uncommon to achieve angular beamwidths up to a fraction of a milliradian with moderate size transmit apertures. Narrow beamwidths allow squeezing of energy into a narrow cone  that leads  to increased transmit energy density and higher signal-to-noise ratio at the receiver. Narrow beamwidths can also help in realizing transfer of energy as well as data simultaneously, a concept known as \emph{simultaneous lightwave information and power transfer} (SLIPT) \cite{Ericsson, PowerLight, Diamantoulakis:TGCN:18, Diamantoulakis:GLOCOM:17, Diamantoulakis:GLOCOM:18, Bashir:TWC:22}. Moreover, narrow beamwidths of laser signals minimize interference with neighboring terminals and endow FSO terminals with security and anti-jamming advantages unavailable to conventional RF systems \cite{Boluda-Ruiz:oe:19}. 
 
 However, communications with narrow beams is not easy,  more so in the case of mobile FSO terminals. FSO terminals need to first acquire the narrow beam laser signal (a process known as \emph{acquisition}) \cite{Bashir:TWC:20, Bashir:TCOMM:21} before the link is setup. When acquisition is achieved successfully, the mobile terminals need to maintain link alignment by tracking the angle-of-arrival of the incoming beam. Therefore, acquisition and tracking subsystems (comprising of hardware as well as algorithms) form important blocks of any modern optical receiver, and much effort has been devoted in recent past to improving the acquisition and tracking performance of FSO communication systems \cite{Kaymak:COMST:18}. 

 Majority of the lasers emit beams with a Gaussian intensity profile \cite{Paschotta:SPIE:08}. In this case, the laser's optical resonator is said to be operating in the \emph{fundamental transverse mode} or $\text{TEM}_{00}$ mode. The Gaussian beam preserves its profile at any point along the beam axis \cite{pampaloni:arXiv:04}. Moreover, the Gaussian profile is also preserved as the beam passes through lens \cite{mandel_wolf:Cambridge:1995, pampaloni:arXiv:04}. Also, single-mode fibers emit laser beams whose profile is closely approximated by a Gaussian distribution. All these attributes render Gaussian beams important and the most widely used intensity model in free-space optical communications.

 \subsection{Motivation of Current Study}
 \begin{figure}
     \centering
     \includegraphics[scale=0.65]{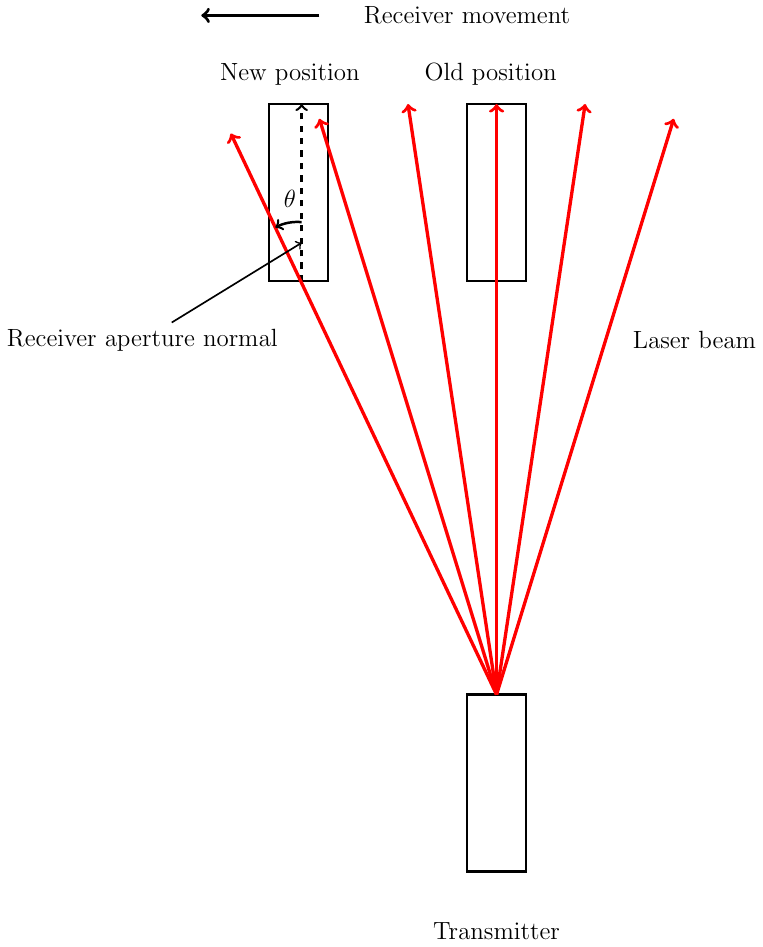}
     \caption{Change in angle-of-arrival with receiver movement.}
     \label{AoA_Tracking}
 \end{figure}
Due to narrow beamwidth of laser beams, free-space optical terminals have to track the incoming beam's angle-of-arrival continuously to maintain sufficient signal-to-noise ratio at the receiving terminal. This fact is illustrated in Fig.~\ref{AoA_Tracking} where the receiving terminal is aligned with the transmitter in the old position (the receiver aperture normal is at the same angle as the incoming beam). In this position, the receiver receives maximum signal energy since the Gaussian beam packs the highest energy density at the center of the beam.  However, when the terminal moves to the new position,  the receive aperture normal makes an angle $|\theta| > 0$ with the incoming beam. We define $\theta$ as the \emph{angle-of-arrival} of the laser beam with respect to the direction of the normal to receiver aperture. In order to maintain the same energy as before, the receiver needs to track this nonzero angle-of-arrival and transmit this information over to the transmitter so that the transmitter can point to the new  receiver location in order to maximize the signal-to-noise ratio. 

\begin{figure}
     \centering
     \includegraphics[scale=0.4]{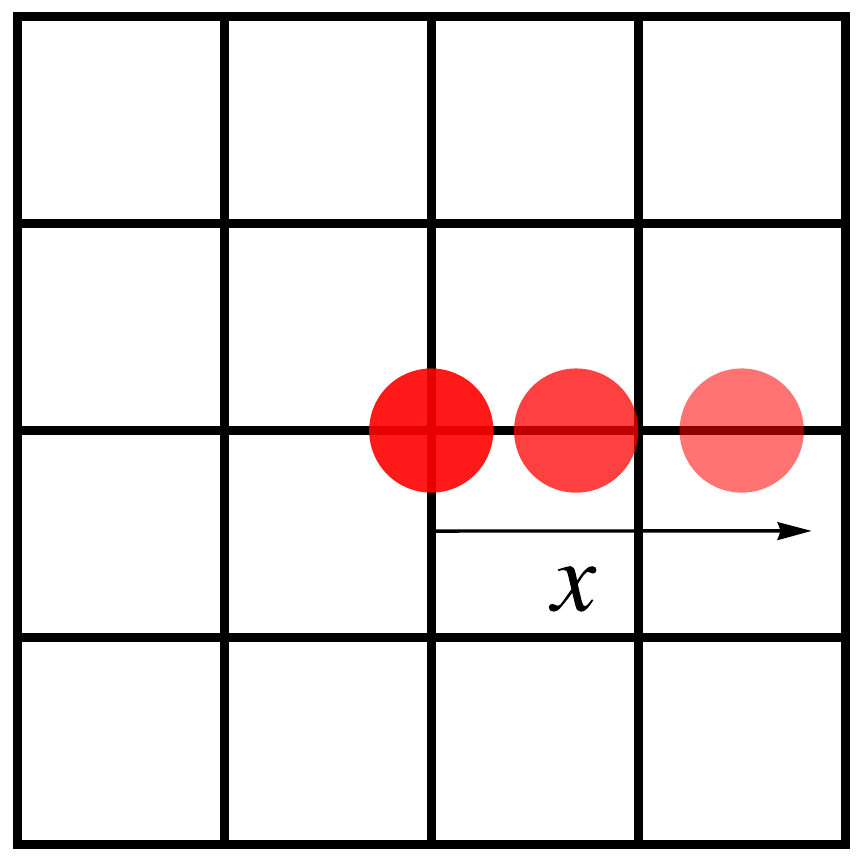}
     \caption{The spot energy decays as it moves away from the center of the detector array. }
     \label{fig21}
 \end{figure}

A number of studies on laser beam tracking \cite{Safi:LCOMM:21, Bashir:TAES:16, Bashir:TAES:17, Bashir:TAES:18, Bashir:OJCOMS:21, Kaymak:COMST:18} indicate that the angle-of-arrival of an incoming laser beam is estimated or tracked based on the location of focused spot on a quad detector or an array of detectors in the focal plane. In a one-dimensional plane, the incoming beam's angle-of-arrival  causes a deviation of the spot position from the center of the array according to the relationship \cite{Goodman:Fourier_Optics:05}
 \begin{align}
     d = F\sin(\theta), 
 \end{align}
where $d$ is the (one-dimensional) Euclidean distance from the center of the array, $F$ is the focal length and $\theta$ is  the angle-of-arrival. In this scenario, if we estimate the deviation $d$, we can infer angle-of-arrival $\theta$. However,  as shown in Fig.~\ref{fig21}, the information of angle-of-arrival is also captured in the energy or brightness of the spot on focal plane. A non-zero angle-of-arrival causes not only the deviation in position of spot but also causes the energy of the spot to diminish. This phenomenon is captured by the dark red spot closer to center of the array and the light red spot further away from the center. The variation in spot energy as a function of angle-of-arrival can be exploited further to enhance the tracking performance of the receiver. We note that the variation of spot energy as a function of angle-of-arrival $\theta$ is highly dependent on the angular beamwdith (denoted by $\phi$) of the laser beam: A highly narrow beam (small $\phi$) leads to sharp variations of energy even for small values of angle-of-arrival. 

The state-of-the-art conventional tracking systems take into account only one parameter for the estimation of angle-of-arrival $\theta$: the spot location on the focal plane array. However, since the spot energy also contains information about the angle-of-arrival, a tracking system that takes into account both the spot location as well spot energy as observations will yield a better tracking performance compared to a system that only rely on spot location as their observations. We expect that the narrower the beam is, the greater the difference in the tracking performance of the two systems.

\subsection{Literature Review}
In this section, we briefly discuss a number of technical papers on pointing, acquisition and tracking (PAT) for FSO applications. A comprehensive survey on PAT systems for FSO---especially from a hardware perspective---is carried out in \cite{Kaymak:COMST:18}.  The studies \cite{Snyder1991, Slocumb:SPIE:90, Bashir:TAES:16, Bashir:TAES:17, Bashir:TAES:18, Bashir:OJCOMS:21} approach tracking from a systems perspective and  infer the angle-of-arrival  based on spot location estimation in focal plane. In their studies, they considered a Poisson channel based on photon counting detectors for  deep space communications. In contrast, the articles \cite{Safi:LCOMM:21, Li:Sensors:19} consider beam tracking based on position sensing algorithms with the help of a quadrant detector in a Gaussian channel. 

Concerning pointing errors, the authors in \cite{Farid:JLT:07} optimize the outage capacity of an optical link in presence of pointing errors, whereas the article \cite{Mai:ao:18} considers adaptive beam control techniques to mitigate the effect of pointing error. Bekkali et al. \cite{Bekkali:JLT:22} devised intelligent lens-based optical-beam-stabilization (OBS) by employing miniature and cost-effective 3-axis voice-coil motors to minimize pointing error for a reliable fiber-to-fiber FSO link. The authors in \cite{Bashir:TCOMM:22} mitigate the effect of pointing error by optimizing the locations of hovering unmanned aerial vehicles (UAV) in a serial FSO link. For acquisition of FSO terminals, the article \cite{Bashir:TWC:20} discusses optimization problems and \cite{Bashir:TCOMM:21} proposes adaptive schemes for enhancing the acquisition performance of FSO terminals. Heyou et al. \cite{liu:arXiv:23} maximize the acquisition performance of a lidar-assisted mobile FSO platform by optimal allocation of power between lidar and optical transmitter assemblies at the ground station.

For readers interested in communications with an array of detectors in free-space optical communications, references \cite{Bashir:TAES:20, Vilnrotter:JLT:04, Vilnrotter:TCOMM:02, Srinivasan:SPIE:16, Bashir:OJCOMS:21} provide comprehensive introduction on the subject. The authors in \cite{Bashir:TAES:19} propose time synchronization algorithms of pulse position modulation symbols based on an array of detectors. In another study, the authors \cite{Bashir:OJCOMS:20} discuss multiple-input-single-output FSO communications based on an array of detectors. Tsai et al. \cite{Tsai:OJCOMS:22} analyze the diversity techniques based on the outputs of detectors of an array receiver to maximize system performance. The study \cite{Bashir:TCOMM:a:21} considers optimal power allocation between beam tracking and symbol detection channels in order to minimize the bit error rate of the optical channel.

\subsection{Contributions of This Study}
 The state-of-the-art tracking systems considered in literature review operate on the information furnished by location of the focal spot to infer the angle-of-arrival. However, in the current study, we have argued that the spot energy is also a function of the angle-of-arrival, and estimation algorithms operating on the  spot energy as well as spot location would yield superior performance compared to algorithms that only take spot location into account. To substantiate our argument, we have derived Cram\'er-Rao lower bounds of angle-of-arrival $\theta$ for quite a few scenarios, and we were able to show that the Cram\'er-Rao bounds based on both spot location as well as energy are lower in magnitude compared to Cram\'er-Rao bounds based exclusively on spot location  for all values of $\theta.$ We also infer from this study that for the same beam power, a smaller beamwidth will lead to more information (about $\theta$) in spot energy and a greater improvement in performance compared to wider beamwidths for values of $\theta$ close to zero. However, as $\theta$ becomes large, narrow beamwidth signals will lose energy rather quickly compared to wider beamwidths, and in order to estimate large $\theta$, wider beamwidth laser signals will outperform signals with narrow beamwidth. In the latter part of the paper, we analyze Cram\'er-Rao bounds for optical channels suffering from pointing error where the (Gaussian distributed) pointing error is treated an additional source of noise in the receiver. 

 This study will find significant application in mobile platform communications where the effect of atmospheric turbulence (on signal energy) is not significant. These include satellite to satellite communications in space,  communications between high altitude platforms in stratosphere, short distance optical wireless channels such as vehicle to vehicle channels, and indoor optical wireless communications based on infrared spectrum. Almost all of these types of channels will find widescale applications in 6G and beyond wireless communications standards \cite{Huawei:6G}. We believe that the results of our study will lead to an efficient laser beam tracking system for mobile FSO terminals that will help realize tens of Tbps level speeds with super-thin narrow Gaussian beams.

\subsection{Paper Organization}
This paper is organized as follows. Section~\ref{system} introduces the system model wherein we derive the relationship between Gaussian beam energy and the angle-of-arrival. Section~\ref{CRLB} deals with the derivation of Cram\'er-Rao lower bound (CRLB) of the angle-of-arrival $\theta$. In this section, we consider the CRLB for two scenarios: i) For the specific scenario where the channel suffers only from Gaussian noise, and ii) the more general case that includes both Gaussian noise and pointing error. We highlight and discuss experimental results in section \ref{simulations}. Finally, we conclude the work in Section~\ref{conclusion} along with a discussion on the implications of the current work and some directions for future work.

\section{Relationship Between Beam Energy and Angle-of-Arrival }\label{system}

As highlighted before,  a major goal of this study is to estimate the angle-of-arrival of a Gaussian beam at the receiver. As a first (important) step, we establish the variation of Gaussian beam energy at the receiver as a function of angle-of-arrival. To achieve this, we first  define Gaussian beam intensity as a function of distance from the beam center through the following relationship:  

 	 	\begin{figure}
		\centering
  	     \includegraphics[scale=0.5]{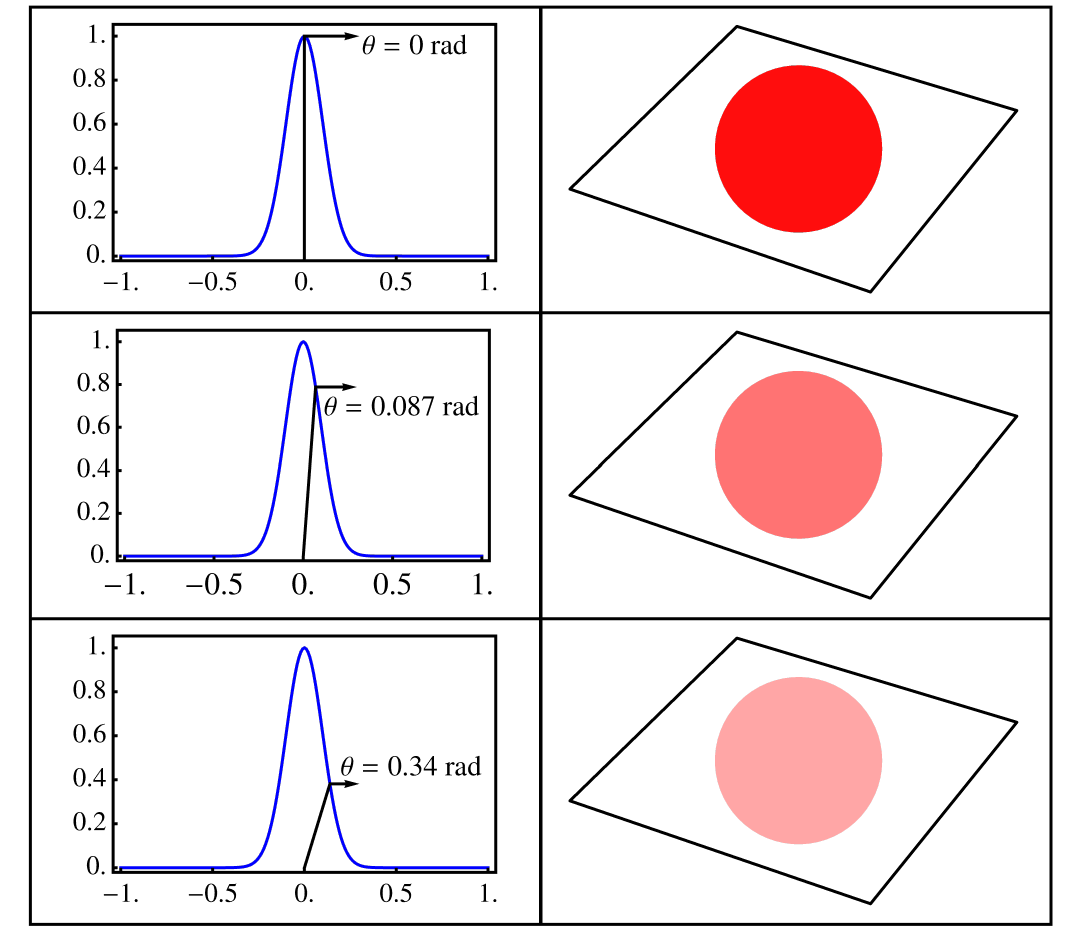}
		\includegraphics[scale=0.475
]{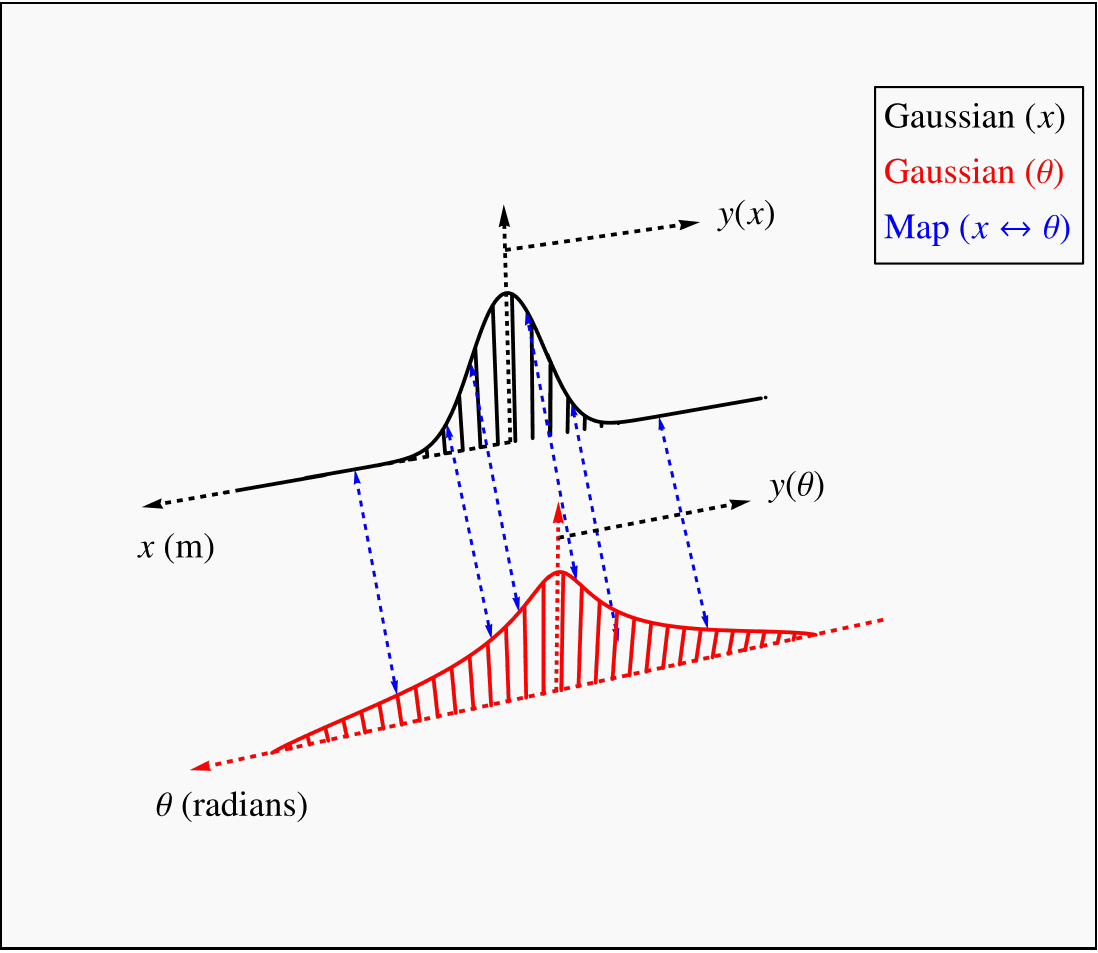}
		\caption{An illustration of power variation in Gaussian beam with respect to the angle-of-arrival $\theta$. The top figure represents the beam intensity variation with respect to angle $\theta$. The bottom figure shows the intensity variation translation from the Gaussian (as a function of distance $x$) to the non-Gaussian distribution (with respect to angle $\theta$). }
		\label{function_mapping}
\end{figure}

\begin{align}
    y(x) = \frac{I_0}{\sqrt{2 \pi  w(L)^2}} \exp \left(-\frac{x^2}{2 w^2(L) }\right),
\end{align}
where $y$ denotes the intensity of Gaussian beam in Watts/m$^2$. Here, we have assumed---without loss of generality---that the beam center lies at the origin. The quantity $I_0$ is the total power in the transmitted beam measured in Watts. The factor $x$ represents the (one-dimensional) distance from the beam center in a plane perpendicular to direction of beam propagation. For a Gaussian beam, the beam radius $w(L)$ evolves with distance $L$ according to the relationship \cite{Trichili:JOSAB:20}
 \begin{align}
     w(L) = w_0 \sqrt{ 1 + \left( \frac{\lambda L}{\pi w_0^2} \right)^2},
 \end{align}
 where $\lambda$ is the wavelength of light signal and $w_0$ is known as the \emph{beam waist}. When $L$ is large, we have that
 \begin{align}
     w(L) &= w_0 \sqrt{ 1 + \left( \frac{\lambda L}{\pi w_0^2} \right)^2} \approx w_0 \sqrt{  \left( \frac{\lambda L}{\pi w_0^2} \right)^2} \nonumber \\
     & = \frac{\lambda}{\pi w_0} L. \label{eq3}
 \end{align}
In \eqref{eq3}, the quantity $\dfrac{\lambda}{\pi w_0}$ is termed as the half-angle \emph{beamwidth}. For the sake of compactness, we will use the notation 
\begin{align}
\phi \coloneqq \dfrac{\lambda}{\pi w_0},
\end{align}
to denote the angular beamwdith. The expression of Gaussian beam in terms of $\phi$ is reproduced below as
\begin{align}
y(x)=\frac{I_0}{\sqrt{2 \pi  (L \phi )^2}} \exp \left(-\frac{x^2}{2 (L \phi )^2}\right). \label{tri_gaussian}
\end{align}

 \begin{figure}
    \centering
    \includegraphics[scale=0.55]{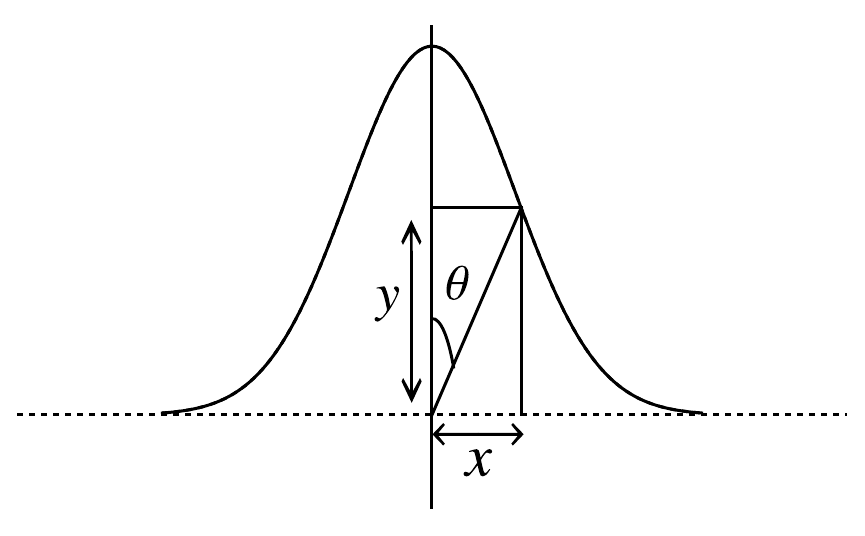}
    \caption{The relationship between Gaussian beam amplitude $y$, the angle-of-arrival $\theta$, and deviation $x$ along $x$-axis.}
    \label{fig11}
\end{figure}
 As shown in Fig.~\ref{fig11}, the Gaussian intensity $y$ as a function of angle-of-arrival $\theta$ and deviation along $x$-axis is 
 \begin{align}
     y = \frac{x}{\tan(\theta)}.
 \end{align}

By inserting $x=y\tan(\theta)$ into \eqref{tri_gaussian}, we have that 
	\begin{align}
          &y   {=}\frac{I_0}{\sqrt{2 \pi  (L \phi )^2}} \exp \left(-\frac{y^2 \tan ^2(\theta )}{2 (L \phi )^2}\right),
        \\
       \Rightarrow & y \exp \left(\frac{y^2 \tan ^2(\theta )}{2 (L \phi )^2}\right) {=}\frac{I_0}{\sqrt{2 \pi  (L \phi )^2}} .\label{eq1}
       \end{align}
       Squaring both sides of \eqref{eq1} and then multiplying both sides of \eqref{eq1} by $\frac{\tan^2(\theta)}{(L \phi)^2}$, we obtain,
\begin{align}
    \frac{y^2 \tan^2(\theta)}{(L\phi)^2} \exp\left( \frac{y^2 \tan^2(\theta)}{(L\phi)^2} \right) = \frac{I_0^2 \tan^2(\theta)}{2 \pi (L \phi)^4}. \label{eq2}
\end{align}
Here, we first note the property of \emph{Lambert $W$} function that for any functions $f$ and $g$, $f = g \exp(g) \implies g = W(f)$. Applying this property to \eqref{eq2}, we have that
       \begin{align}
       &\frac{y^2 \tan^2(\theta)}{(L\phi)^2} = W\left( \frac{I_0^2 \tan^2(\theta)}{2 \pi (L \phi)^4} \right) \\
    &\implies y{=} \frac{L \phi }{\tan (\theta )} \sqrt{W\left(\frac{I_0^2 \tan ^2(\theta )}{2 \pi  (L \phi )^4}\right)}.\\
    \end{align}
 Finally, we note that
 \begin{align}
     x = y \tan(\theta) \implies x = L \phi  \sqrt{W\left(\frac{I_0^2 \tan ^2(\theta )}{2 \pi  (L \phi )^4}\right)}. \label{x}
 \end{align}
Substituting the expression of $x$ in \eqref{x} into \eqref{tri_gaussian}, we obtain a mathematically tractable expression of intensity $y$ in terms of angle-of-arrival $\theta$ as
\begin{align}
    y = \frac{I_0}{\sqrt{2 \pi  (L \phi )^2}} \exp \left(-\frac{1}{2} W\left(\frac{I_0^2 \tan ^2(\theta )}{2 \pi L^4   \phi ^4}\right)\right) \cdot \mathbbm{1}_{(-\frac{\pi}{2}, \frac{\pi}{2})}(\theta),
\end{align}
where $\mathbbm{1}_A(x)$ represents the \emph{indicator} function that is equal to unity whenever $x \in A$ for any measurable set $A$ and zero otherwise.

\section{Cram\'er-Rao Lower Bound of Angle-of-Arrival $\theta$ }\label{CRLB}
In this section, we derive the Cram\'er-Rao bounds for the angle-of-arrival $\theta$. For the sake of simplicity, we assume a one-dimensional array of detectors in the focal plane where the spot image is captured for angle-of-arrival tracking. In this scenario,  the peak intensity---captured by the receiver telescope lens and projected onto a focal plane array---is given by the expression
\begin{align}
    \Lambda_0(\theta) &=\frac{I_0}{\sqrt{2 \pi  (L \phi )^2}} \exp \left(-\frac{1}{2} W\left(\frac{I_0^2 \tan ^2(\theta )}{2 \pi L^4   \phi ^4}\right)\right) \pi a^2 ,\label{lambda_0}
\end{align}
where $\phi$ is the beamwidth in radians, $L$ is the link distance in meters, $I_0$ is the peak (received signal) intensity impinging on the receive aperture and $a$ is the receive telescope radius in meters. Here, we have assumed that the beam footprint $\pi (L\phi)^2 \gg \pi a^2$. 
The signal intensity captured on the focal plane array is modeled by a (one-dimensional) Gaussian distribution as
\begin{align}
   \Lambda_s = \frac{\Lambda_0}{\sqrt{2 \pi \rho^2}} \exp\left(-\frac{(x-x_0)^2 }{2\rho^2} \right) \cdot \mathbbm{1}_{\mA}(x), \label{lambda_s}
\end{align}
where $\mA$ is the region of the square array, $\rho$ is the spot size on the focal plane and $x_0$ is the location of spot center on the one-dimensional array. Here, the center of the array is fixed at origin.

The deviation $x_0$ of the spot center location on the array is related to the angle-of-arrival as 
\begin{align}
    x_0(\theta) = F \sin (\theta) \label{x_0},
\end{align}
where $F$ is the focal length of receiver telescope lens in meters. We note that $\Lambda_s$ in \eqref{lambda_s} is a function of angle-of-arrival $\theta$ due to dependence on  the energy $\Lambda_0(\theta)$ of the spot as well as the location $x_0(\theta)$ of spot center on the focal plane array.

The output of thermal-noise (modeled by Gaussian distribution) limited array is 
\begin{align}
    \bm{Y} = \bm{\Lambda} + \bm{X},
\end{align}
where $\bm{Y}$ is $M \times 1$ output vector, $\bm{\Lambda}$ is $M \times 1$ signal vector and $\bm{X}$ is an $M \times 1$ noise vector. Here, 
\begin{align}
\bm{Y}& = \begin{bmatrix}
        Y_0 & Y_1 & \dotsm & Y_{M-1}
    \end{bmatrix}^T, \\
    \bm{\Lambda}& = \begin{bmatrix}
        \Lambda_0 & \Lambda_1 & \dotsm & \Lambda_{M-1}
    \end{bmatrix}^T, \\
    \bm{X}& = \begin{bmatrix}
        X_0 & X_1 & \dotsm & X_{M-1}
    \end{bmatrix}^T,
\end{align}
where $Y_m$ is the output of the $m$th element of the array:
\begin{align}
    Y_m = \Lambda_m + X_m, \quad m=0, 1, \dotsc, M-1. \label{output}
\end{align}
In \eqref{output}, \begin{align}
\Lambda_m& \coloneqq \int_{A_m} \Lambda_s \, dx \label{lambda_m} \\
&= \frac{\Lambda_0}{\sqrt{2\pi \rho^2}} \int_{A_m} \exp\left( - \frac{(x-x_0)^2}{2 \rho^2} \right) \, dx \label{Lambda_m}
\end{align}
is the signal output of the $m$th detector of the array.  Here, $A_m$ is the region of the $m$th detector. The factor $N_m$ represents a Gaussian random variable that models thermal noise of the $m$th detector. Here, $X_m \sim \mathcal{N}(0, \sigma_n^2)$ for all $m$ and $X_m \perp X_j$ for positive integers $m$ and $j$ such that $m \neq j, 0 \leq m, j \leq M-1$. The quantity $\sigma_n^2$ is the thermal noise variance. 

The log-likelihood of a single observation is described by
\begin{align}
    p(Y_m | \theta) &= \frac{1}{\sqrt{2 \pi  \sigma_n^2 }} \exp \left( - \frac{\left(Y_m - \Lambda_m \right)^2}{2 \sigma_n^2} \right), \\
    \ln p(Y_m|\theta)&=-\ln(\sqrt{2 \pi  \sigma_n^2 }) -  \frac{\left(Y_m - \Lambda_m \right)^2}{2 \sigma_n^2}, \label{loglikelihood} \\ 
    \ln p(\bm{Y}|\theta)& = -M\ln(\sqrt{2 \pi  \sigma_n^2 }) -  \sum_{m=0}^{M-1}\frac{\left(Y_m - \Lambda_m \right)^2}{2 \sigma_n^2}.
\end{align}

Taking the second derivative of the log-likelihood function with respect to $\theta$, we obtain
\begin{align}
    \frac{\partial^2 \ln p (\bm{Y}|\theta)}{\partial \theta^2} &= \sum_{m=0}^{M-1} \left( \frac{\Lambda_m''\left( Y_m - \Lambda_m \right)}{\sigma_n^2} - \frac{\left( \Lambda_m' \right)^2}{\sigma_n^2}  \right),
    \end{align}
    and the Fisher information of $\theta$ in random vector $\bm{Y}$ is
    \begin{align}
    -\E \left[\frac{\partial^2 \ln p (\bm{Y}|\theta)}{\partial \theta^2}\right] & = \frac{1}{\sigma_n^2}\sum_{m=0}^{M-1}  \left( \Lambda_m' \right)^2.  \label{FI}
\end{align}

We now analyze the Fisher information of $\theta$ given in \eqref{FI}. After a few steps, it can be shown that $\Lambda_m'$ (see the expression of $\Lambda_m$ in \eqref{Lambda_m}) can be decomposed into two terms $\alpha_m$ and $\beta_m$ as
\begin{align}
\Lambda_m' = \alpha_m + \beta_m, \label{lambda_m_prime}
\end{align}
where
\begin{align}
   & \alpha_m \coloneqq \frac{\Lambda_0'}{\sqrt{2\pi \rho^2}} \int_{A_m} \exp\left( - \frac{(x-x_0)^2}{2\rho^2} \right) \, dx, \\
    & \beta_m \coloneqq \frac{\Lambda_0 x_0'}{\sqrt{2\pi}\rho^3}\int_{A_m} \exp\left( - \frac{(x-x_0)^2}{2\rho^2} \right) (x-x_0)\, dx. 
\end{align}
Here, we highlight that the factor $\beta_m$ corresponds to the information of $\theta$ provided by spot location in the focal plane, whereas $\alpha_m$ corresponds to the information provided by spot energy.  


After a number of mathematical manipulations, the derivative of $\Lambda_0$, as a function of $\theta$, is shown to be
\begin{align}
    \Lambda_0'
    =-\frac{I_0 (\pi a^2) \csc (\theta ) \sec (\theta ) e^{-\frac{1}{2} W\left(\frac{I_0^2 \tan ^2(\theta )}{2 L^4 \phi ^4}\right)} W\left(\frac{I_0^2 \tan ^2(\theta )}{2 L^4 \phi ^4}\right)}{\sqrt{2\pi} \sqrt{L^2 \phi ^2} \left(W\left(\frac{I_0^2 \tan ^2(\theta )}{2 L^4 \phi ^4}\right)+1\right)}. \label{lambda_0_prime}
\end{align}
Moreover, 
\begin{align}
    x_0' = F \cos(\theta).\label{x_0_prime}
\end{align}

Finally, by substituting expression of $\Lambda_0'$ and $x_0'$ into \eqref{lambda_m_prime}, we obtain the Fisher information of $\theta$ from \eqref{FI}. The final form of the Fisher information (in terms of $\alpha_m$ and $\beta_m$) is

    \begin{align}
    &-\E \left[\frac{\partial^2 \ln p (\bm{Y}|\theta)}{\partial \theta^2}\right]  = \frac{1}{\sigma_n^2}\sum_{m=0}^{M-1}  \left( \alpha_m + \beta_m \right)^2 \nonumber \\
    & = \frac{1}{\sigma_n^2}\sum_{m=0}^{M-1}  \beta_m^2+\frac{1}{\sigma_n^2}\sum_{m=0}^{M-1}   \alpha_m^2  + \frac{1}{\sigma_n^2}\sum_{m=0}^{M-1}  2 \alpha_m \beta_m. \label{FI1}
\end{align}
In \eqref{FI1}, the first summation term on the right hand side (containing $\beta_m$'s) corresponds to the Fisher information of $\theta$ based solely on the spot location in the focal plane, whereas the second and third summation terms comprise additional Fisher information when spot energy is considered as an extra  observation in addition to spot location.

The Cram\'er-Rao lower bound (denoted by \textsf{CRLB}) of $\theta$ can be reached by taking the inverse of Fisher information:
\begin{align}
    \textsf{CRLB}(\theta)&= \frac{\sigma_n^2}{\sum_{m=0}^{M-1}(\alpha_m+\beta_m)^2 }.
\end{align}

\section{Cram\'er-Rao Bounds for General Receiver (Thermal Noise and Pointing Error) }
For the general receiver that suffers both from thermal noise and pointing error, the energy captured by receiver lens fluctuates as
\begin{align}
    \Lambda_0(\theta+\Theta_p)= \frac{I_0 (\pi a^2)}{\sqrt{2 \pi  (L \phi )^2}} \exp \left(-\frac{1}{2} W\left(\frac{I_0^2 \tan ^2(\theta+\Theta_p)}{2 \pi L^4   \phi ^4}\right)\right)  ,\label{Lambda_0_pointing}
\end{align}
where $\Theta_p$ represents the \emph{angular pointing error}. We assume that the pointing error is distributed as a Gaussian random variable: $\Theta_p \sim \mathcal{N}(0, \sigma^2_p)$, where  $\sigma_p^2$ captures the angular pointing error variance. In addition to random fluctuation in spot energy $\Lambda_0$, the angular pointing error also causes random fluctuations in spot center position on the array  according to the relationship
\begin{align}
     x_0(\theta+\Theta_p) = F \sin(\theta + \Theta_p). \label{x_0}
\end{align}
When pointing error variance $\sigma^2_p$ is close to zero, we have an approximate result for \eqref{x_0} based on the fact that $\cos(z) \approx 1$ and $\sin(z) \approx z$ for small $z$. The approximate result is,
\begin{align}
    x_0(\theta +\Theta_p)& \approx F \sin(\theta) + F\cos(\theta) \Theta_p. \label{x_01}
\end{align}

If the pointing error is small, then an approximate relationship for spot energy in \eqref{Lambda_0_pointing} also holds through Taylor Series expansion:
\begin{align}
    \Lambda_0(\theta + \Theta_p) \approx \Lambda_0(\theta) + \Lambda_0'\Theta_p. \label{X_p}
\end{align}
  We note that the approximation in \eqref{X_p} gets better as the pointing error variance $\sigma_p^2$ approaches zero.

Under angular pointing error, the final expression for the intensity---captured on the focal plane---becomes 
\begin{align}
   &\Lambda_s = \frac{\Lambda_0(\theta+\Theta_p)}{\sqrt{2 \pi \rho^2}} \exp\left(-\frac{(x-x_0(\theta+\Theta_p))^2 }{2\rho^2} \right) \cdot \mathbbm{1}_{\mA}(x) \label{lambda_s_1} \\
   & \approx \left(\frac{\Lambda_0(\theta)}{\sqrt{2 \pi \rho^2}}+ \frac{\Lambda_0' \Theta_p }{\sqrt{2 \pi \rho^2}} \right) \nonumber \\
   & \times \exp\left(-\frac{(x-F \sin(\theta) - F\cos(\theta) \Theta_p )^2 }{2\rho^2} \right) \cdot \mathbbm{1}_{\mA}(x). \label{eq5}
\end{align}

It is  shown in the appendix that for $|\Theta_p|$ small, $\Lambda_s$ can be written (approximately) as the sum of a signal term and a noise term:
\begin{align}
    &\Lambda_s \approx \frac{\Lambda_0(\theta)}{\sqrt{2 \pi \rho^2}} \exp\left(-\frac{(x-F \sin(\theta))^2}{2\rho^2}\right)  \nonumber \\
    & + \frac{1}{\sqrt{2\pi \rho^2}}\exp\left(-\frac{(x-F \sin(\theta))^2}{2\rho^2}\right) \nonumber \\
    & \times \left( \Lambda_0(\theta) \frac{(x-F\sin(\theta))F\cos(\theta)}{\rho^2} + \Lambda_0' \right) \Theta_p. \label{Lambdas}
\end{align}
In \eqref{Lambdas}, the first term in the sum on the right hand side is the signal component and the second term in the sum (containing $\Theta_p$) is the noise component. This fact  implies that the ``signal'' output of the $m$th element of the array is made of a signal term, $\mathcal{S}_m$, and a (pointing error induced) noise term, $\mathcal{X}_m$ as
\begin{align}
    &\Lambda_m \! = \! \int_{A_m} \Lambda_s \, dx \approx \mathcal{S}_m + \mathcal{X}_m ,
    \end{align}
    where 
    \begin{align}
   & \mathcal{S}_m \coloneqq \int_{A_m} \frac{\Lambda_0(\theta)}{\sqrt{2 \pi \rho^2}} \exp\left(-\frac{(x-F \sin(\theta))^2}{2\rho^2}\right) dx, \nonumber \\
    & \mathcal{X}_m \coloneqq   \int_{A_m} \frac{1}{\sqrt{2\pi \rho^2}}\exp\left(-\frac{(x-F \sin(\theta))^2}{2\rho^2}\right) \nonumber \\
    & \times \left( \Lambda_0(\theta) \frac{(x-F\sin(\theta))F\cos(\theta)}{\rho^2} + \Lambda_0' \right) dx \, \Theta_p. \label{S_m_X_m}
\end{align}
Thus, the output of the $m$th detector of the array is given by 
\begin{align}
    Y_m &= \Lambda_m + X_m \nonumber \\
    &\approx \mathcal{S}_m + Z_m 
\end{align}
where $Z_m$ is the total noise that is the sum of two independent Gaussian random variables: $X_m$ which represents thermal noise with variance $\sigma_n^2$ and $\mathcal{X}_m$ which represents noise due to pointing error. The variance of $\mathcal{X}_m$ is $\gamma_m^2 \sigma_p^2$. The constant $\gamma_m$ is defined as the integral
\begin{align}
    \gamma_m(\theta) & \coloneqq \int_{A_m} \frac{1}{\sqrt{2\pi \rho^2}}\exp\left(-\frac{(x-F \sin(\theta))^2}{2\rho^2}\right) \nonumber \\
    & \times \left( \Lambda_0(\theta) \frac{(x-F\sin(\theta))F\cos(\theta)}{\rho^2} + \Lambda_0' \right) dx.
\end{align}
The variance of $Z_m$, denoted by $\sigma_m^2$, is 
\begin{align}
\sigma_m^2(\theta) \coloneqq \gamma_m^2(\theta) \sigma_p^2+\sigma_n^2. \label{eq8}
\end{align}

Note: Here, we assume that the angle-of-arrival $\theta$ is sampled (or updated) at a frequency much slower than the frequency  of the random process driving the pointing error $\Theta_p$. The frequency at which $\theta$ is updated depends on the speed of the mobile terminal---and for most practical scenarios---the update happens on the order of a few hundreds of milliseconds. In contrast, the pointing error varies on the order of one millisecond or less. Therefore, it is safe to assume that the Gaussian noise resulting from pointing error is white for the sampling rates at which the angle-of-arrival $\theta$ is updated. 

The log-likelihood function for the general receiver is given by the expression
\begin{align}
    &\ln p(\bm{Y}|\theta) = -\sum_{m=0}^{M-1}\left(\ln\left(\sqrt{2 \pi  \sigma_m^2 }\right) + \frac{\left(Y_m - \mathcal{S}_m \right)^2}{2 \sigma_m^2}\right) \nonumber \\
    & = -\sum_{m=0}^{M-1}\left(\ln\left(\sqrt{2 \pi  (\gamma_m^2(\theta) \sigma_p^2+\sigma_n^2) }\right) + \frac{\left(Y_m - \mathcal{S}_m(\theta) \right)^2}{2 (\gamma_m^2(\theta) \sigma_p^2+\sigma_n^2) }\right).
\end{align}
It can be shown that the Fisher information of $\theta$ is
\begin{align}
    -\mathbbm{E} \left[\frac{\partial^2 \ln p(\bm{Y}|\theta)}{\partial \theta^2} \right] = \sum_{m=0}^{M-1} \left(\frac{(\mathcal{S}_m')^2}{\sigma_m^2} + \left( \frac{\sigma_m'}{\sigma_m} \right)' - \left( \frac{\sigma_m'}{\sigma_m^3} \right)' \sigma_m^2 \right),
\end{align}
where, through \eqref{eq8},  $\sigma_m= \sqrt{\gamma_m^2(\theta) \sigma_p^2+\sigma_n^2}.$

\section{Experimental Results And Discussion}
\label{simulations}

In this section, we analyze and comment on the CRLB plots of angle-of-arrival $\theta$ for various channel conditions. Since the CRLB is an even function of angle-of-arrival $\theta$, we only consider the positive $\theta$ axis for Fig.~\ref{result2_fig} through Fig.~\ref{result6_fig}.  The default parameter values for the plots are shown in  Table~\ref{table_1}.
\begin{table}[H]
\centering
\scalebox{1.1}{\begin{tabular}{|c|c|c|} \hline
{Notation} & {Name / Unit} & {Default Values}                               \\ \hline
                      $I_0$        &  received peak power                                   & 1 mW to 100 mW                                      \\ \hline
                      
                      $\sigma_n$        &       thermal noise standard deviation                       & $10^{-6}$ Volts                                  \\ \hline
                      
                    $\theta$          &   angle-of-arrival                              & $-\pi/2$ to $\pi/2$ rads                                                           \\ \hline
                     
                     $\phi$         &    angular beamwidth                                 & 1 to 10 mrad                                                                                                                      \\ \hline
                     $L$         &    link distance    &                          100 to 1000 meters                                                                                                                      \\ 
                     \hline 
                     $F$ & focal length & 1 mm \\
                     \hline 
                     $|\mathcal{A}|$ & array area & 4 mm$^2$ \\ \hline 
                     $\rho$ & spot radius & 0.2 mm \\
                     
                     \hline
\end{tabular}}
\caption{ Default values of system parameters. } \label{table_1}
\end{table}

	 	\begin{figure}
        
		\centering
		\includegraphics[scale=0.6]{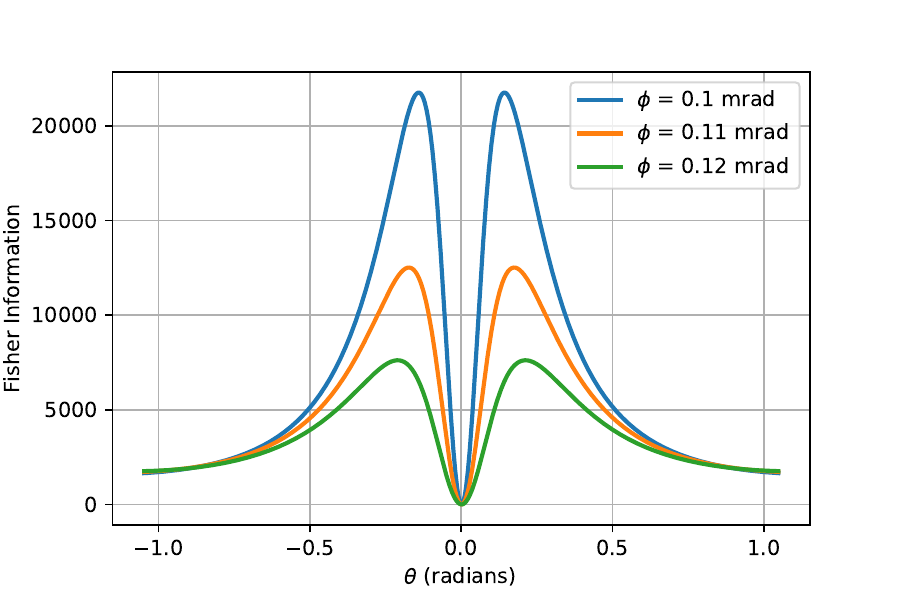}
		\caption{Fisher information with respect to angle-of-arrival $\theta$ for different beamwidth $\phi$.}
   \label{result1_fig}
		\end{figure}
  
 Fig.~\ref{result1_fig} represents the Fisher information of $\theta$ based solely on the variation of spot energy on the focal plane. This Fisher information is defined by \eqref{FI}, and is plotted in Fig.~\ref{result1_fig} as function of angle-of-arrival $\theta$. We note that the Fisher information is maximized at the value of $\theta$ that corresponds to maximum rate of change of energy in the Gaussian beam as a function of $\theta$. We also note that the Fisher information is zero at $\theta = 0$. This is because of the fact that the  slope of Gaussian beam is zero when $\theta = 0$, and any small change in $\theta$ near zero does not lead to any significant changes in beam energy that can be utilized for estimating the angle-of-arrival with sufficient accuracy. In other words, the beam energy delivers zero information about the angle-of-arrival $\theta$ when $\theta = 0$.
 
 	 	\begin{figure}
		\centering
		\includegraphics[scale=0.6]{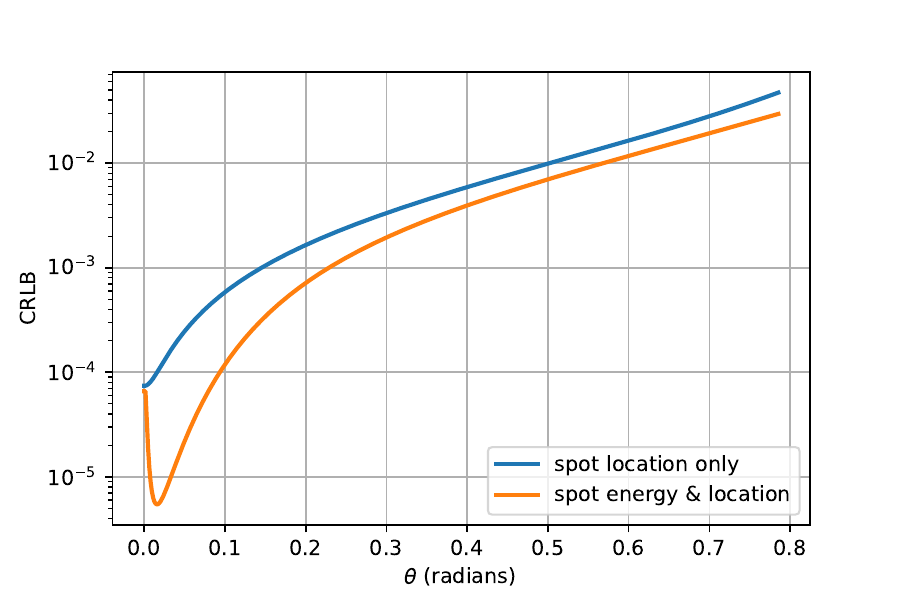}
		\caption{Cram\'er-Rao lower bound comparison.}
\label{result2_fig}
		\end{figure}

  Fig.~\ref{result2_fig} represents the Cram\'er-Rao lower bound (CRLB) (inverse of Fisher information) curves as a function of $\theta$. In this figure, we also compare the CRLB for two scenarios: i) Only spot location information is used to infer $\theta$, and ii) both spot energy and location are used to estimate $\theta$. We note that the spot energy provides significant amount of information in addition to spot location that helps improve (lowers) the CRLB.

 	 	\begin{figure}
		\centering
		\includegraphics[scale=0.6]{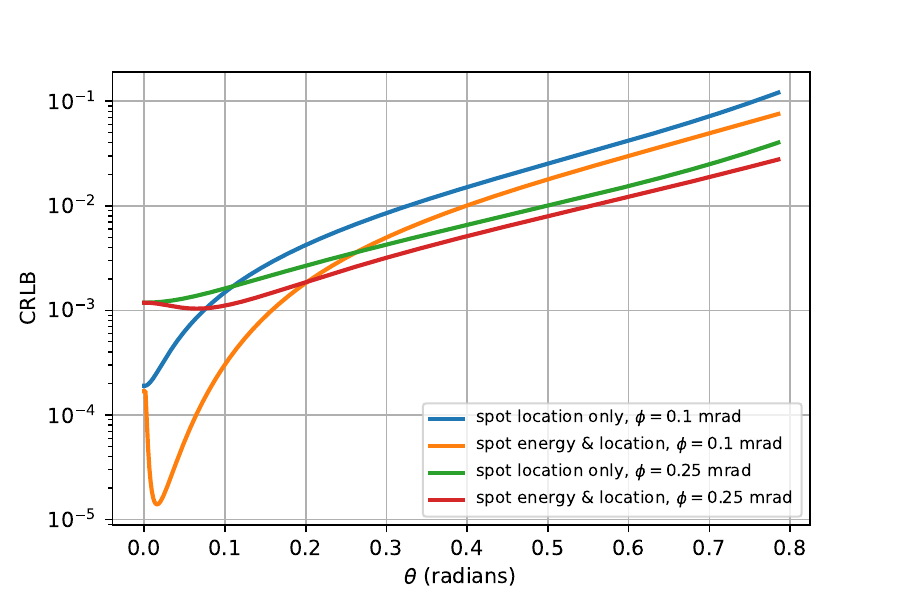}
		\caption{Cram\'er-Rao lower bound comparison for different beamwidth $\phi$. }
 \label{result3_fig}
		\end{figure}

 Fig.~\ref{result3_fig} depicts the CRLB curves as a function of $\theta$ for different beamwidth $\phi$ with the assumption that the energy in the beam is constant for different beamwidth values. We observe that a narrower beam leads to a smaller CRLB for smaller values of $\theta$. However, for large values of angle-of-arrival $\theta$, the energy of a narrow beam decays at a faster rate than a beam with a larger beamwidth, and the CRLB of a narrower beam exceeds that of a wider beam beyond a certain value of $\theta$. Thus, depending on how quickly the angle-of-arrival is sampled or updated, we may want to choose the appropriate beamwidth that minimizes the CRLB. For a slow moving terminal or a high sampling rate system, we can estimate the angle-of-arrival more accurately with a narrow beam since $\theta$ is not varying significantly. On the other hand, for a fast moving target or a low sampling rate system, we may want to use a wider beamwidth since the deviation in $\theta$ can be large from one sampling instant to the next. 

 Fig.~\ref{result4_fig} and Fig.~\ref{result5_fig} represent the CRLB of $\theta$ as a function of number of detectors $M$. Here, we assume that the major source of noise in the receiver is due to background radiation which depends only on the active area of the detector array. For a fixed array area, a larger number of detectors yields minimization of CRLB (albeit at the cost of a higher computational complexity). We observe in Fig.~\ref{result4_fig} that the CRLB decreases monotonically as we increase the numbers of detectors. Fig.~\ref{result5_fig} represents the CRLB based on i) spot location alone and ii) spot energy as well as location for two different types of detector arrays; one for the number of detectors $M = 4$ and the other for $M = 16$. For small number of detectors, the CRLB curves exhibit a ``wave like'' effect with peaks and troughs. This is due to the fact that the area of array is constant, and a small number of detectors imply large area per detector. When the area of each detector is large compared to the spot footprint, the CRLB dips when the spot crosses over to the neighboring detector, and reaches a peak when the spot lies in the middle of the detector region \cite{Bashir:OJCOMS:21}.

\begin{figure}
		\centering
		\includegraphics[scale=0.6]{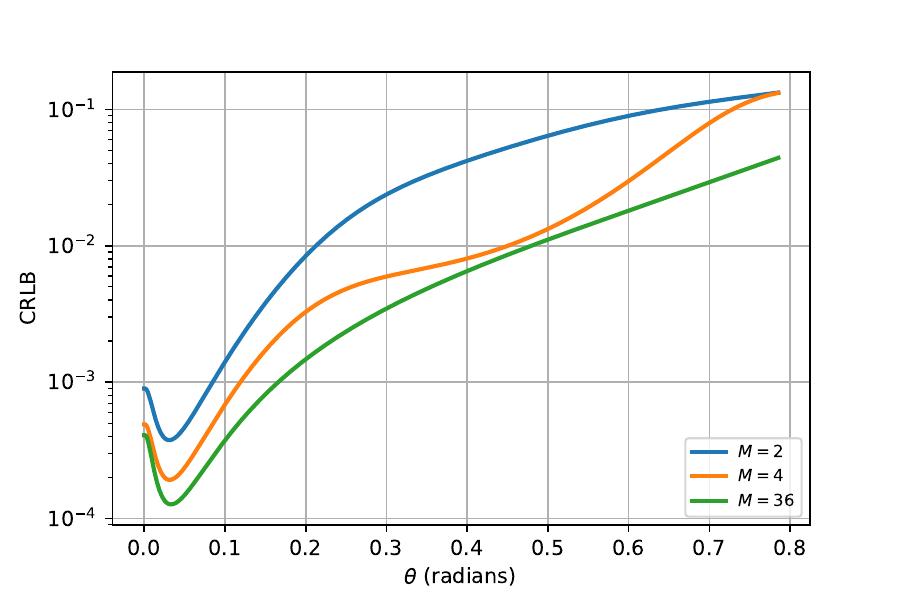}
		\caption{Cram\'er-Rao lower bound comparison for different number of detectors $M$ in the array. }
 \label{result4_fig}
		\end{figure}

\begin{figure}
		\centering
		\includegraphics[scale=0.6]{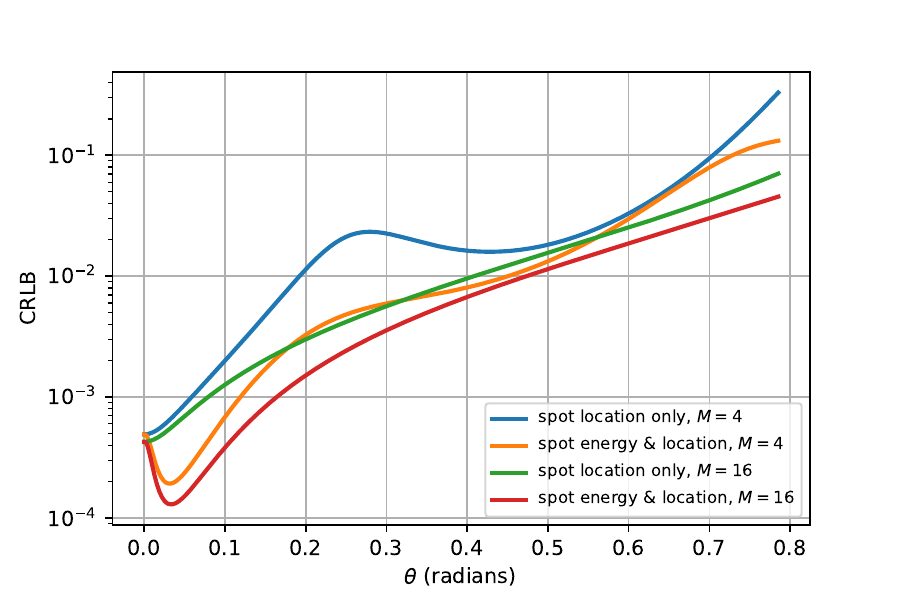}
		\caption{Cram\'er-Rao lower bound comparison for different number of detectors $M$ in the array. }
 \label{result5_fig}
		\end{figure}


Finally, Fig.~\ref{result6_fig} shows the CRLB as a function of $\theta$ for different value of pointing error standard deviation $\sigma_p$. Here, we have assumed that $\sigma_p$ is only a small fraction of the angular beamwdith $\phi$ in order to justify small pointing error approximations made in \eqref{x_01}, \eqref{X_p} and \eqref{S_m_X_m}. The beamwidth in this set of experiments was fixed at 0.2 radians in order to account for pointing error. 

\begin{figure}
		\centering
		\includegraphics[scale=0.6]{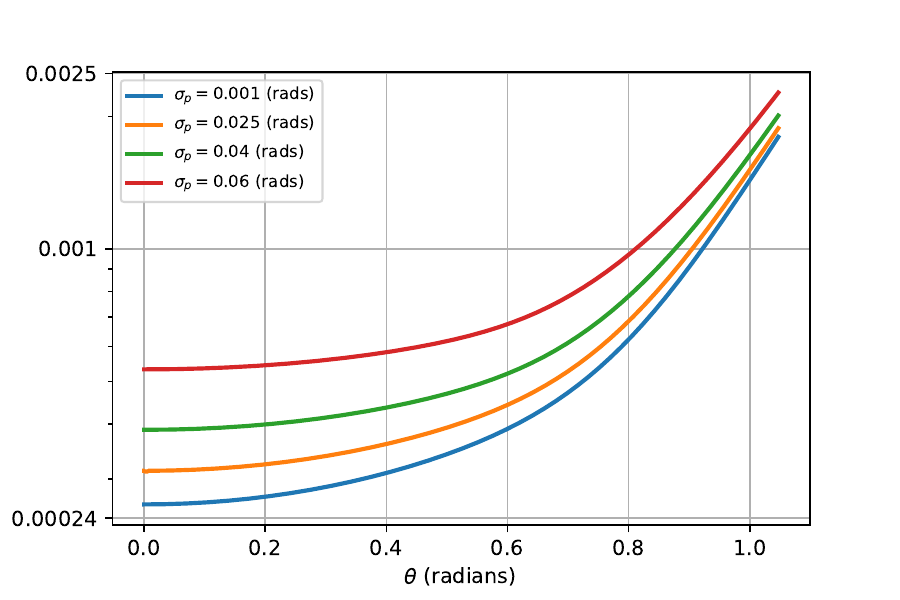}
		\caption{Cram\'er-Rao lower bound comparison for different values of pointing error standard deviation $\sigma_p$.  }
 \label{result6_fig}
		\end{figure}

\section{Conclusion and Future Work} \label{conclusion}

In this study, we proposed that the angle-of-arrival $\theta$ of narrow Gaussian beams can be estimated more accurately if the variation in spot energy as a function of $\theta$ is taken into account as an extra observation in addition to the spot location on the focal plane of an optical receiver.  We first established the relationship between the angle-of-arrival and the energy in Gaussian beam. Thereafter, we showed---through the derived CRLB plots---that significantly better estimation of angle-of-arrival can be achieved if spot energy is incorporated into the estimation process in addition to observations based on spot location. Depending on the beamwidth of the received Gaussian beam, the difference in performance can be significant, especially at smaller values of angle-of-arrival.

As part of future study, we want to study the dynamical problem in which the angle-of-arrival is assumed to  evolve in time as a (Gaussian) random process. For this dynamical model, we consider applying Bayesian filtering algorithms---such as Kalman and particle filters---to track the angle-of-arrival. For the dynamical model, we want to quantify the improvement in tracking performance based on our proposed approach in this paper.

\appendix
The Taylor series expansion of the exponential term in \eqref{eq5} is
\begin{align}
    &\exp\left(-\frac{(x-F \sin(\theta) - F\cos(\theta) \Theta_p ) ^2 }{2\rho^2} \right) \nonumber \\
    &= \exp \! \left(-\frac{(x-F \sin(\theta))^2}{2\rho^2} \right) \exp \!\left( \frac{2 (x-F\sin(\theta)) F\cos(\theta)\Theta_p}{2\rho^2} \right) \nonumber \\
    & \times \exp\left( - \frac{F^2\cos^2(\theta)\Theta^2_p}{2\rho^2} \right). \label{eq6}
\end{align}
For $|\Theta_p| \ll 1$, we ignore the last exponential term in \eqref{eq6}. Under this approximation and by representing the second exponential term through its first order Taylor series approximation, we have that the intensity on the focal plane is
\begin{align}
&\Lambda_s \approx \exp\left(-\frac{(x-F \sin(\theta))^2}{2\rho^2}\right) \left(\frac{\Lambda_0(\theta)}{\sqrt{2 \pi \rho^2}}+ \frac{\Lambda_0' \Theta_p }{\sqrt{2 \pi \rho^2}} \right) \nonumber \\
& \times \left( 1 +  \frac{ (x-F\sin(\theta)) F\cos(\theta)\Theta_p}{\rho^2} \right) \nonumber \\
& = \exp\left(-\frac{(x-F \sin(\theta))^2}{2\rho^2}\right) \nonumber \\
& \times \left( \frac{\Lambda_0(\theta)}{\sqrt{2 \pi \rho^2}} +  \frac{\Lambda_0(\theta)}{\sqrt{2 \pi \rho^2}} \frac{(x-F\sin(\theta)) F\cos(\theta)\Theta_p}{\rho^2}\right. \nonumber \\
& \left. + \frac{\Lambda_0' \Theta_p}{\sqrt{2 \pi \rho^2}} + \frac{\Lambda_0' (x-F\sin(\theta)) F\cos(\theta) \Theta_p^2}{\sqrt{2\pi \rho^6}} \right).
\end{align}
By ignoring the terms containing $\Theta_p^2$, we have that
\begin{align}
    &\Lambda_s \approx \exp\left(-\frac{(x-F \sin(\theta))^2}{2\rho^2}\right) \frac{\Lambda_0(\theta)}{\sqrt{2 \pi \rho^2}} \nonumber \\
    & + \frac{1}{\sqrt{2\pi \rho^2}}\exp\left(-\frac{(x-F \sin(\theta))^2}{2\rho^2}\right) \nonumber \\
    & \times \left( \Lambda_0(\theta) \frac{(x-F\sin(\theta))F\cos(\theta)}{\rho^2} + \Lambda_0' \right) \Theta_p.
\end{align}

\bibliographystyle{IEEEtran}
\bibliography{references.bib}

\end{document}